\documentclass[11pt]{article}

\usepackage[T1]{fontenc}
\usepackage{lmodern}
\usepackage[margin=1in]{geometry}
\usepackage{amsmath,amssymb,amsfonts,amsthm}
\usepackage{bm}
\usepackage{mathtools}
\usepackage{microtype}
\usepackage{natbib}
\usepackage{hyperref}
\usepackage{xurl}

\hypersetup{
  colorlinks=true,
  linkcolor=blue,
  citecolor=blue,
  urlcolor=blue
}

\newtheorem{theorem}{Theorem}[section]
\newtheorem{corollary}[theorem]{Corollary}
\renewcommand{\eqref}[1]{equation~\ref{#1}}

\title{From Scalar Games to Unconstrained Network Zero-Sum Games:\\
Transfer Principles for Gradient-Based Learning}

\author{
Gaoqi Xue\\
\small Department of Industrial and Systems Engineering\\
\small Rensselaer Polytechnic Institute, Troy, NY, USA\\
\small \texttt{xueg@rpi.edu}
\and
James P. Bailey\\
\small Department of Industrial and Systems Engineering\\
\small Rensselaer Polytechnic Institute, Troy, NY, USA\\
\small \texttt{bailej6@rpi.edu}
}
\date{}

\begin{document}
\maketitle

\begin{abstract}
We show that, for a broad class of gradient-based learning rules, scalar zero-sum games form a complete test family for convergence in unconstrained network zero-sum games. Specifically, suppose a learning rule achieves last-iterate or time-average convergence on every scalar game of magnitude $a$ whenever $\eta\leq f(a)$, where $f$ is nonincreasing. Then the same guarantee holds on an arbitrary network zero-sum game with stacked game operator $\mathbf{G}$ whenever $\eta\leq f(\lVert\mathbf{G}\rVert_2)$. More generally, finite-time rate guarantees transfer mode by mode: the network response is bounded by the largest response of its scalar canonical games, and under compatible linear initialization conventions the worst-case finite-time network response is exactly the largest scalar canonical rate. Thus, understanding the learning dynamics on scalar games is sufficient to characterize universal convergence and worst-case rates over the corresponding class of network games. The result follows from the classical planar decomposition of real skew-symmetric game operators. Related mode-wise reasoning has appeared implicitly in spectral analyses of particular learning algorithms and in prior work using these decompositions for game representation and visualization; here we isolate the resulting transfer principle for gradient-based learning. We discuss implications for simplifying convergence analysis, time-average guarantees, low-dimensional geometric arguments, accelerated-method design, and worst-case numerical testing.
\end{abstract}


\section{Introduction}

Zero-sum games are a basic model of adversarial interaction and appear throughout game theory, online learning, robust control, and machine learning~\citep{vonNeumann28,vonNeumannTheoryOfGames,bailey2019fast,isaacs1965,goodfellow2014generative}. Gradient methods for saddle-point problems have a long history, dating at least to the work of Arrow, Hurwicz, and Uzawa~\citep{ArrowHurwicz1958}. Even in unconstrained bilinear games, simple first-order methods can behave very differently. Simultaneous gradient descent--ascent (GDA) may diverge, alternating updates may cycle, optimistic methods can converge, and implicit discretizations can remain stable for much larger learning rates~\citep{Bailey20Regret,daskalakis2018limit,mokhtari2020convergence,xue2026imgd}. These differences have produced a substantial literature on the convergence of learning dynamics in zero-sum games.

Spectral structure has been exploited in several analyses of gradient-based learning in bilinear games. For two-player bilinear games, singular values enter exact convergence conditions and rate calculations for a range of gradient methods~\citep{zhang2020bilinear}. Other analyses use characteristic polynomials, frequency-domain methods, or algorithm-specific eigenvalue arguments~\citep{anagnostides2020robust,demontbrun2025ogda}. Real Schur decomposition has also been used explicitly to reduce particular learning dynamics to planar systems, including implicit midpoint gradient descent~\citep{xue2026imgd}. Together, these works illustrate the usefulness of low-dimensional spectral structure, generally within analyses of particular algorithms rather than as a reusable scalar-to-network transfer principle.

The underlying game decomposition is itself not new. Real skew-symmetric matrices admit a canonical representation in $2\times2$ rotational blocks. In game-theoretic settings, this representation has been used to decompose cyclic competitive structure into planar ``disc games'' and to support visualization, feature extraction, and low-rank approximation~\citep{balduzzi2018reevaluating,strang2024pta}. Recent work further develops these coordinates as a latent space for learning and selection dynamics~\citep{lechon2025disc}. We therefore take the decomposition as known machinery and ask a different question:

\begin{quote}
\emph{When does a convergence theorem proved on scalar zero-sum games automatically become a theorem for arbitrary network zero-sum games?}
\end{quote}

Our main result gives a simple transfer principle for the gradient-based update family considered in this paper. Consider the scalar zero-sum game
\begin{equation}
\max_{u\in\mathbb{R}}\min_{v\in\mathbb{R}} a u v,
\qquad a>0.
\label{eq:scalar-game-intro}
\end{equation}
Suppose a learning rule achieves last-iterate or time-average convergence on every such scalar game whenever
\[
\eta\leq f(a),
\]
where $f$ is nonincreasing. Then the same rule achieves the corresponding convergence guarantee on every unconstrained network zero-sum game whenever
\[
\eta\leq f(\lVert\mathbf{G}\rVert_2),
\]
where $\mathbf{G}$ is the stacked skew-symmetric game operator. More generally, when the admissible scalar parameters are not described by a single monotone threshold, the network parameters must satisfy the scalar conditions associated with every canonical frequency of $\mathbf{G}$. Equivalently, for this update class, scalar bilinear games form a complete test family for universal last-iterate and time-average convergence guarantees on unconstrained network zero-sum games.

The transfer is particularly natural for last-iterate convergence because many existing proofs are already spectral. For time-average convergence, however, the scalar proof need not be spectral: it may instead use a finite equilibrium residual, a geometric trajectory argument, or another direct analysis. Our theorem separates that algorithm-specific scalar proof from the dimension-lifting step. Once a scalar argument establishes the desired time-average guarantee, the same low-dimensional reduction transfers it to the network game.

Our contribution is therefore methodological rather than algorithmic. We do not propose a new learning rule or a new decomposition. Instead, we isolate a reusable transfer principle that makes precise when low-dimensional analysis already suffices for an arbitrary network game. The exact results in this paper concern unconstrained bilinear network zero-sum games and gradient-based rules whose stacked dynamics share the same temporal coefficients. Affine linear costs are covered by translation, and we comment after the main theorem on limited forms of algorithmic asymmetry. Fully asymmetric learning rules and constrained simplex dynamics remain nontrivial extensions.

The paper is organized as follows. Section~\ref{sec:prelim} introduces network zero-sum polymatrix games, the gradient-based update family, and the known canonical decomposition. Section~\ref{sec:transfer} establishes preservation of Nash equilibria and the scalar-to-network transfer results. Section~\ref{sec:useful} discusses where the transfer principle is useful: simplifying and tightening convergence proofs, visualizing low-dimensional dynamics, motivating acceleration through canonical modes, and implications for low-dimensional numerical testing. Section~\ref{sec:conclusion} concludes.

\section{Preliminaries}
\label{sec:prelim}

Throughout, bold lowercase letters denote vectors and bold uppercase letters denote matrices; scalar quantities are unbolded.

\subsection{Network zero-sum polymatrix games}

Consider a network $\mathcal{N}=(\mathcal{V},\mathcal{E})$ with agents
$i\in\mathcal{V}=\{1,\ldots,n\}$. Agent $i$ selects an unconstrained action
$\mathbf{x}_i\in\mathbb{R}^{d_i}$. For each interacting pair
$\{i,j\}\in\mathcal{E}$, let
$\mathbf{A}^{(ij)}\in\mathbb{R}^{d_i\times d_j}$ describe the effect of
agent $j$ on agent $i$. The payoff to agent $i$ is
\begin{equation}
    u_i(\mathbf{x})
    =
    \sum_{j:\{i,j\}\in\mathcal{E}}
    \mathbf{x}_i^\top \mathbf{A}^{(ij)}\mathbf{x}_j.
    \label{eq:network-payoff}
\end{equation}
We assume each pairwise interaction is zero-sum:
\begin{equation}
    \mathbf{A}^{(ij)}
    =
    -\big(\mathbf{A}^{(ji)}\big)^\top.
    \label{eq:pairwise-zerosum}
\end{equation}
Indeed, the contribution of the pair $\{i,j\}$ to their combined payoff is
\[
    \mathbf{x}_i^\top\mathbf{A}^{(ij)}\mathbf{x}_j
    +
    \mathbf{x}_j^\top\mathbf{A}^{(ji)}\mathbf{x}_i
    =0,
\]
so $\sum_i u_i(\mathbf{x})=0$ for every joint action.

Let
\[
    \mathbf{x}
    =
    \begin{bmatrix}
        \mathbf{x}_1^\top & \cdots & \mathbf{x}_n^\top
    \end{bmatrix}^\top
\]
denote the stacked action vector, and define the stacked game operator
\begin{equation}
\mathbf{G}=
\begin{bmatrix}
\mathbf{0} & \mathbf{A}^{(12)} & \cdots & \mathbf{A}^{(1n)}\\
\mathbf{A}^{(21)} & \mathbf{0} & \cdots & \mathbf{A}^{(2n)}\\
\vdots & \vdots & \ddots & \vdots\\
\mathbf{A}^{(n1)} & \mathbf{A}^{(n2)} & \cdots & \mathbf{0}
\end{bmatrix},
\label{eq:network-operator}
\end{equation}
where a zero block is used whenever $\{i,j\}\notin\mathcal{E}$. By
\eqref{eq:pairwise-zerosum}, $\mathbf{G}^\top=-\mathbf{G}$. Moreover,
\[
    \nabla_{\mathbf{x}_i}u_i(\mathbf{x})
    =
    \sum_{j:\{i,j\}\in\mathcal{E}}
    \mathbf{A}^{(ij)}\mathbf{x}_j,
\]
so the stacked gradient field is
\begin{equation}
    F(\mathbf{x})=\mathbf{G}\mathbf{x}.
    \label{eq:game-field}
\end{equation}

The familiar two-player bilinear zero-sum game is recovered as
\[
    \max_{\mathbf{x}_1}\min_{\mathbf{x}_2}
    \mathbf{x}_1^\top\mathbf{A}\mathbf{x}_2,
\]
with
\begin{equation}
\mathbf{G}=
\begin{bmatrix}
\mathbf{0}&\mathbf{A}\\
-\mathbf{A}^\top&\mathbf{0}
\end{bmatrix}.
\label{eq:twoplayer-G}
\end{equation}
For the unconstrained bilinear games considered here, a joint action
$\mathbf{x}^\star$ is a Nash equilibrium if and only if
\begin{equation}
    \mathbf{G}\mathbf{x}^\star=0.
    \label{eq:nash-null}
\end{equation}
Thus the Nash set is the nullspace of the game operator.

\paragraph{Linear costs and nonzero equilibria.}
We state the main results using the homogeneous field
$F(\mathbf{x})=\mathbf{G}\mathbf{x}$ for notational simplicity, but the same
analysis applies to affine game fields
\begin{equation}
    F(\mathbf{x})=\mathbf{G}\mathbf{x}+\mathbf{b},
    \label{eq:affine-field}
\end{equation}
which arise naturally when bilinear games include linear costs
\citep{kim2025parallelizable,bailey20261,xue2026imgd,zhang2020bilinear}.
Suppose an equilibrium $\mathbf{x}^\star$ exists, so that
$\mathbf{G}\mathbf{x}^\star+\mathbf{b}=0$. Under the translated coordinates
$\mathbf{z}=\mathbf{x}-\mathbf{x}^\star$,
\[
    F(\mathbf{x})=\mathbf{G}\mathbf{z}.
\]
For the update family below, the same translation removes the affine term
whenever $\sum_{s\in\mathcal H}\alpha_s=1$: after substituting
$\mathbf{x}^t=\mathbf{z}^t+\mathbf{x}^\star$, all constant terms cancel.
Hence the decomposition and transfer results extend unchanged to the affine
setting whenever a Nash equilibrium exists. Existence is equivalent to
$-\mathbf{b}\in\operatorname{range}(\mathbf{G})$.

\subsection{Gradient-based update rules}

We consider finite-memory gradient-based rules of the form
\begin{align}
    \mathbf{x}_i^t
    =
    \sum_{s\in \mathcal{H}_i}
    \left(
        \alpha_s \mathbf{x}_i^{t-s}
        +
        \lambda_s
        \sum_{j\neq i}
        \mathbf{A}^{(ij)}\mathbf{x}_j^{t-s}
    \right).
    \tag{Gradient-Based Update Rule}
    \label{eq:update}
\end{align}
Our transfer results assume that all agents use the same update rule, i.e., that they use common coefficients $\alpha_s$ and $\lambda_s$ across agents. Under this algorithmic symmetry assumption, the stacked dynamics can be written as
\begin{equation}
    \mathbf{x}^t
    =
    \sum_{s\in\mathcal H}
    \left(
        \alpha_s \mathbf{I}+
        \lambda_s \mathbf{G}
    \right)\mathbf{x}^{t-s}.
    \label{eq:stacked}
\end{equation}
This common-coefficient structure allows the same orthogonal transformation to decouple both the game and the learning dynamics into canonical scalar games.
This family includes several standard rules. For example, simultaneous GDA and optimistic GDA are respectively
\[
\mathbf{x}^t=\mathbf{x}^{t-1}+\eta\mathbf{G}\mathbf{x}^{t-1}
\]
and
\[
\mathbf{x}^t=\mathbf{x}^{t-1}+2\eta\mathbf{G}\mathbf{x}^{t-1}-\eta\mathbf{G}\mathbf{x}^{t-2}.
\]
Using the standard duplicated/time-shifted stacked representation, two-player alternating gradient descent can be written as
\[
\mathbf{x}^t=\mathbf{x}^{t-2}+\eta\mathbf{G}\mathbf{x}^{t-1},
\]
which is again of the form \eqref{eq:stacked}; this representation and its multi-agent network extension are used in the analysis of alternating gradient descent in network zero-sum games~\citep{bailey20261}. The implicit rule
\[
\mathbf{x}^t=\mathbf{x}^{t-1}+\eta\mathbf{G}\mathbf{x}^{t}
\]
is obtained by allowing the current gradient in the history. These examples illustrate that \eqref{eq:update} covers explicit, optimistic, alternating, and implicit gradient-based dynamics while retaining a common linear representation.

When $\sum_{s\in\mathcal H}\alpha_s=1$ and $\sum_{s\in\mathcal H}\lambda_s\neq0$, the fixed points of \eqref{eq:stacked} are exactly the Nash equilibria: the fixed-point condition reduces to $\mathbf{G}\mathbf{x}^\star=0$. For a finite-memory rule requiring stored history, a standard initialization convention is to set every stored iterate equal to the starting point, e.g., $\mathbf{x}^{-s}=\mathbf{x}^0$ for each required pre-history index $s$. The decomposition and transfer arguments below do not depend on this particular convention: any fixed initialization rule can be transformed consistently into the canonical coordinates and inherited by each scalar subsystem. We only require that the zero-frequency recurrence remain fixed at its initialized value; this holds under the standard initialization whenever $\sum_{s\in\mathcal H}\alpha_s=1$. Convergence statements below are therefore naturally interpreted as convergence to the Nash set.

\subsection{Canonical decomposition of the game operator}

Because $\mathbf{G}$ is real and skew-symmetric, there exists an orthogonal matrix $\mathbf{Q}$ such that
\begin{equation}
\mathbf{G}=\mathbf{Q}\mathbf{J}\mathbf{Q}^\top,
\label{eq:schur}
\end{equation}
where $\mathbf{J}$ is block diagonal with nonzero blocks
\begin{equation}
\mathbf{J}(\omega_j)=
\begin{bmatrix}
0&\omega_j\\
-\omega_j&0
\end{bmatrix},
\qquad j=1,\ldots,\ell,
\label{eq:block}
\end{equation}
and a possible zero block~\citep{golu2013matrix}. The positive numbers $\omega_j$ are the magnitudes of the nonzero imaginary eigenvalues of $\mathbf{G}$. In the two-player representation \eqref{eq:twoplayer-G}, they are the nonzero singular values of $\mathbf{A}$.

This decomposition is classical and has appeared in several game-theoretic settings. \citet{balduzzi2018reevaluating} use the same planar structure to represent cyclic evaluation matrices, while \citet{strang2024pta} develop it into Principal Trade-off Analysis (PTA), interpreting the blocks as planar disc games. \citet{lechon2025disc} further use related coordinates as a latent space for learning dynamics. Our contribution is not the matrix decomposition itself; we use it as the mechanism behind the transfer results in Section~\ref{sec:transfer}.

Let
\[
\mathbf{y}^t=\mathbf{Q}^\top\mathbf{x}^t.
\]
Substituting \eqref{eq:schur} into \eqref{eq:stacked} gives
\begin{equation}
\mathbf{y}^t
=
\sum_{s\in\mathcal H}
\left(\alpha_s\mathbf{I}+\lambda_s\mathbf{J}\right)\mathbf{y}^{t-s}.
\label{eq:transformed}
\end{equation}
Hence each pair $(y_{2j-1},y_{2j})$ evolves according to
\begin{align}
 y_{2j-1}^t
 &=\sum_{s\in\mathcal H}
 \left(\alpha_s y_{2j-1}^{t-s}+\lambda_s\omega_j y_{2j}^{t-s}\right),\nonumber\\
 y_{2j}^t
 &=\sum_{s\in\mathcal H}
 \left(\alpha_s y_{2j}^{t-s}-\lambda_s\omega_j y_{2j-1}^{t-s}\right).
\label{eq:mode-dynamics}
\end{align}
These are exactly the same gradient-based update rules applied to the scalar zero-sum games
\begin{equation}
\max_u\min_v \omega_j u v.
\label{eq:canonical-game}
\end{equation}
The remaining coordinates correspond to zero frequencies and evolve according to the recurrence induced by the $\alpha_s$ terms alone.

\section{Transferability of Low-Dimensional Results}
\label{sec:transfer}

\subsection{Preservation of Nash equilibria}

Before transferring convergence guarantees, we first verify that the low-dimensional representation preserves the target of the dynamics. This question is natural whenever a game is replaced by simpler component games, and it is immediate here from the nullspace characterization of Nash equilibria.

\begin{theorem}[Nash preservation]
\label{thm:nash-preservation}
A point $\mathbf{x}^\star$ is a Nash equilibrium of the original network zero-sum game if and only if $\mathbf{y}^\star=\mathbf{Q}^\top\mathbf{x}^\star$ is a Nash equilibrium of the transformed game. Equivalently,
\[
\mathbf{G}\mathbf{x}^\star=0
\qquad\Longleftrightarrow\qquad
\mathbf{J}\mathbf{y}^\star=0.
\]
\end{theorem}

\begin{proof}
Since $\mathbf{G}=\mathbf{Q}\mathbf{J}\mathbf{Q}^\top$ and $\mathbf{Q}$ is orthogonal,
\[
\mathbf{G}\mathbf{x}^\star=0
\iff
\mathbf{Q}^\top\mathbf{G}\mathbf{x}^\star=0
\iff
\mathbf{J}\mathbf{Q}^\top\mathbf{x}^\star=0.
\]
Setting $\mathbf{y}^\star=\mathbf{Q}^\top\mathbf{x}^\star$ gives the result.
\end{proof}

For the gradient-based update family, Theorem~\ref{thm:nash-preservation} also preserves fixed points whenever $\sum_{s\in\mathcal H}\alpha_s=1$ and $\sum_{s\in\mathcal H}\lambda_s\neq0$, because the fixed-point condition reduces to $\mathbf{G}\mathbf{x}^\star=0$ in the original coordinates and to $\mathbf{J}\mathbf{y}^\star=0$ after transformation.

\subsection{Scalar convergence and rate guarantees transfer to network games}

The central result is that both convergence conditions and quantitative rates can be established on the scalar games \eqref{eq:canonical-game} and then lifted to the network game. Fix any initialization convention for the stored history of a finite-memory rule. Let $r_T^{\mathrm{LI}}(\eta,a)$ denote the smallest constant such that
\begin{equation}
    \|\mathbf{z}^T-\mathbf{z}^\star\|
    \leq
    r_T^{\mathrm{LI}}(\eta,a)
    \|\mathbf{z}^0-\mathbf{z}^\star\|
    \label{eq:r-li-def}
\end{equation}
for every initial point in the scalar game of magnitude $a$ under that convention. Define $r_T^{\mathrm{TA}}(\eta,a)$ analogously with the time-averaged iterate $\bar{\mathbf{z}}^T$ in place of $\mathbf{z}^T$. When a statement applies to either notion, we write $r_T(\eta,a)$. We define the corresponding network response $R_T(\eta,\mathbf{G})$ analogously as the smallest constant for which the same relative-error bound holds for every network initialization, with the comparison Nash equilibrium chosen to agree with the initialized zero-frequency coordinates. The transfer upper bound below applies to any fixed initialization convention inherited consistently by the scalar subsystems; repeated-history initialization is only one common choice. For the exact worst-case identity below, we call an initialization convention \emph{compatible and linear} if the stored history depends linearly on the initial error and can be applied independently on each canonical plane; repeated-history initialization satisfies this condition.

\begin{theorem}[Scalar-to-network convergence and rate transfer]
\label{thm:transfer}
Consider any \eqref{eq:update} parameterized by a learning rate $\eta$. Suppose there exists a nonincreasing function $f$ such that, for every scalar zero-sum game
\[
    \max_u\min_v a u v,
    \qquad a>0,
\]
the rule has a last-iterate rate $r_T^{\mathrm{LI}}(\eta,a)$ whenever $\eta\leq f(a)$. Then, for every unconstrained network zero-sum game with positive canonical frequencies $\omega_1,\ldots,\omega_\ell$,
\begin{equation}
    \|\mathbf{x}^T-\mathbf{x}^\star\|
    \leq
    \max_{1\leq j\leq\ell} r_T^{\mathrm{LI}}(\eta,\omega_j)
    \|\mathbf{x}^0-\mathbf{x}^\star\|
    \label{eq:rate-transfer}
\end{equation}
whenever
\begin{equation}
    \eta\leq f(\|\mathbf{G}\|_2),
    \label{eq:transfer-lr}
\end{equation}
where $\mathbf{x}^\star$ agrees with the initialized zero-frequency coordinates. In particular, if $r_T^{\mathrm{LI}}(\eta,a)\to0$ for every $a>0$, then the network dynamics converge to the Nash set.

The same statement holds for time-average rates after replacing $\mathbf{x}^T$ by $\bar{\mathbf{x}}^T$ and $r_T^{\mathrm{LI}}$ by $r_T^{\mathrm{TA}}$.
\end{theorem}

\begin{proof}
Let $\omega_1,\ldots,\omega_\ell$ be the positive frequencies in the canonical decomposition of $\mathbf{G}$. Under the orthogonal transformation $\mathbf{y}^t=\mathbf{Q}^\top\mathbf{x}^t$, the network dynamics reduce to the same update rule applied to the scalar games with magnitudes $\omega_1,\ldots,\omega_\ell$.

Because $f$ is nonincreasing and
\[
    \omega_j\leq \max_k\omega_k=\|\mathbf{G}\|_2,
\]
the condition $\eta\leq f(\|\mathbf{G}\|_2)$ implies $\eta\leq f(\omega_j)$ for every $j$. Hence the assumed scalar rate guarantee applies to every canonical game.

Let $\mathbf{z}_j^T$ and $\mathbf{z}_j^\star$ denote the two-dimensional coordinates associated with frequency $\omega_j$. Choose $\mathbf{x}^\star$ so that its zero-frequency coordinates agree with their initialized values. Because $\mathbf{Q}$ is orthogonal, Euclidean distance is preserved and the transformed error decomposes across the canonical planes:
\[
    \|\mathbf{x}^T-\mathbf{x}^\star\|^2
    =
    \|\mathbf{Q}^\top(\mathbf{x}^T-\mathbf{x}^\star)\|^2
    =
    \sum_{j=1}^{\ell}
    \|\mathbf{z}_j^T-\mathbf{z}_j^\star\|^2.
\]
Applying the scalar rate bound on each canonical plane gives
\begin{align*}
    \|\mathbf{x}^T-\mathbf{x}^\star\|^2
    &\leq
    \sum_{j=1}^{\ell}
    r_T^{\mathrm{LI}}(\eta,\omega_j)^2
    \|\mathbf{z}_j^0-\mathbf{z}_j^\star\|^2\\
    &\leq
    \left(\max_{1\leq j\leq\ell}r_T^{\mathrm{LI}}(\eta,\omega_j)\right)^2
    \|\mathbf{x}^0-\mathbf{x}^\star\|^2.
\end{align*}
Taking square roots yields \eqref{eq:rate-transfer}. If $r_T^{\mathrm{LI}}(\eta,\omega_j)\to0$ for every canonical frequency, then the right-hand side converges to zero.

For time-average convergence, $\bar{\mathbf{y}}^T=\mathbf{Q}^\top\bar{\mathbf{x}}^T$ because the coordinate transformation is linear. The same argument therefore applies with $\bar{\mathbf{x}}^T$ and $r_T^{\mathrm{TA}}$.
\end{proof}

\paragraph{Algorithmic asymmetry.}
The common-coefficient assumption is sufficient for the orthogonal transfer result above, but some limited forms of asymmetry can still be accommodated. For example, fixed positive player-specific scalar learning-rate multipliers can be absorbed through a block-diagonal rescaling, after which the effective game operator remains skew-symmetric. The resulting convergence transfer holds in the rescaled coordinates, although exact Euclidean rate identities in the original coordinates may acquire conditioning factors. More general asymmetric update rules, in which agents use different histories or update formulas, require separate analysis.

\begin{corollary}[Exact worst-mode response]
\label{cor:worst-mode}
Under a compatible linear initialization convention, the worst-case finite-time network response satisfies
\begin{equation}
    R_T(\eta,\mathbf{G})
    =
    \max_{1\leq j\leq\ell} r_T(\eta,\omega_j).
    \label{eq:exact-worst-mode}
\end{equation}
The identity holds for both last-iterate and time-average response.
\end{corollary}

\begin{proof}
Theorem~\ref{thm:transfer} gives the upper bound. For the reverse inequality, let $j^\star\in\arg\max_j r_T(\eta,\omega_j)$ and choose an initial condition whose error lies entirely in the corresponding canonical plane and attains the scalar worst-case response. All other canonical modes remain at equilibrium, so the resulting network response equals $r_T(\eta,\omega_{j^\star})$. Hence the upper bound is attained.
\end{proof}

\begin{corollary}[Scalar games form a complete test family]
\label{cor:complete-test}
Fix a learning rate $\eta$ and a set of positive frequencies $\Omega$. A gradient-based rule in the class considered here has universal last-iterate (respectively, time-average) convergence on every unconstrained network zero-sum game whose positive canonical frequencies lie in $\Omega$ if and only if it has the corresponding convergence property on every scalar game $\max_u\min_v a u v$ with $a\in\Omega$. Under a compatible linear initialization convention, the worst-case finite-time response over this network class is
\[
    \sup_{a\in\Omega} r_T(\eta,a).
\]
\end{corollary}

\begin{proof}
If the scalar property holds for every $a\in\Omega$, Theorem~\ref{thm:transfer} applies to every canonical frequency of every network in the class, and Corollary~\ref{cor:worst-mode} gives the rate statement. Conversely, each scalar game is itself a two-agent unconstrained network zero-sum game, so any universal network guarantee must also hold on every scalar game in $\Omega$.
\end{proof}

The monotone-threshold form in Theorem~\ref{thm:transfer} is chosen because it matches many learning-rate theorems. More generally, if the scalar rule is admissible for a set of parameter values that depends on $a$, then the network rule must satisfy the scalar condition associated with every canonical frequency $\omega_j$. Thus the admissible network parameters are obtained by intersecting the admissible scalar regions.

Theorem~\ref{thm:transfer} makes explicit a principle that already appears implicitly in several last-iterate analyses. Spectral and singular-value arguments often reduce a method to separate frequencies and identify the most restrictive one~\citep{zhang2020bilinear,anagnostides2020robust,demontbrun2025ogda,xue2026imgd}. Our purpose is not to claim that mode-wise reasoning is new, but to isolate the reusable transfer statement: once the scalar theorem is established for the update family considered here, a separate high-dimensional convergence proof is unnecessary.

\paragraph{Last-iterate and time-average guarantees.}
Theorem~\ref{thm:transfer} deliberately treats last-iterate and time-average guarantees in parallel. The difference lies in how the scalar premise is established, not in the transfer step. Last-iterate analyses often expose the canonical frequencies directly through characteristic roots or singular values. Time-average analyses may instead use geometric or optimization-style quantities that are more natural in low dimensions.

For the unconstrained games considered here, the usual unrestricted Nash gap is generally infinite away from equilibrium. A simple finite alternative is the \emph{Nash residual}
\begin{equation}
    \mathcal{R}_{\mathbf{G}}(\mathbf{x})
    :=\|\mathbf{G}\mathbf{x}\|_2,
    \label{eq:nash-residual}
\end{equation}
which vanishes exactly on the Nash set (and becomes $\|\mathbf{G}\mathbf{x}+\mathbf{b}\|_2$ in the affine formulation \eqref{eq:affine-field}). The residual is preserved by the orthogonal transformation:
\[
    \mathcal{R}_{\mathbf{G}}(\mathbf{x})
    =\|\mathbf{J}\mathbf{Q}^\top\mathbf{x}\|_2.
\]
Consequently, a scalar argument establishing $\|\mathbf{J}(a)\bar{\mathbf{z}}^T\|_2\to0$ for every scalar game transfers by the same blockwise proof as Theorem~\ref{thm:transfer}. More generally, the theorem is agnostic to whether the scalar time-average guarantee is proved by a residual, a direct trajectory calculation, or another finite certificate. This is one reason the reduction can be useful even when the original proof is not spectral.

\section{Where the Transfer Principle Is Useful}
\label{sec:useful}

The transfer theorem is useful when it changes how a learning rule is analyzed, designed, or tested. We illustrate several such uses below.

\subsection{Simplifying and tightening convergence analysis}
\label{sec:simplifying}

\paragraph{Optimistic gradient descent--ascent.}
For OGD,
\[
\mathbf{x}^t=\mathbf{x}^{t-1}+2\eta\mathbf{G}\mathbf{x}^{t-1}-\eta\mathbf{G}\mathbf{x}^{t-2}.
\]
On a scalar game of magnitude $\omega$, the dynamics depend on the game only through $q=\eta\omega$. The scalar characteristic roots lie strictly inside the unit disk exactly when
\[
|q|<\frac{1}{\sqrt{3}}.
\]
Hence Theorem~\ref{thm:transfer} immediately gives
\[
\eta<\frac{1}{\sqrt{3}\,\lVert\mathbf{G}\rVert_2}
\]
for the network game. The sharp scalar threshold and related optimal-rate analyses are known~\citep{zhang2020bilinear,anagnostides2020robust,demontbrun2025ogda}; the transfer theorem isolates the scalar calculation as the only algorithm-specific step.

\paragraph{Implicit midpoint gradient descent.}
IMGD applies the implicit midpoint rule to the game field:
\[
\mathbf{x}^t=\mathbf{x}^{t-1}+\frac{\eta}{2}\mathbf{G}\left(\mathbf{x}^t+\mathbf{x}^{t-1}\right).
\]
On a scalar game,
\[
\left(\mathbf{I}-\frac{\eta}{2}\mathbf{J}(\omega)\right)\mathbf{z}^t
=
\left(\mathbf{I}+\frac{\eta}{2}\mathbf{J}(\omega)\right)\mathbf{z}^{t-1}.
\]
Because $\mathbf{J}(\omega)$ is skew-symmetric, the one-step map is orthogonal for every real $\eta$. This establishes bounded last iterates for every learning rate. Separately, the full IMGD analysis proves an $O(1/T)$ time-average convergence guarantee~\citep{xue2026imgd}. Both scalar facts can then be lifted to the network game by the transfer principle.

\paragraph{A negative result: simultaneous GDA.}
On a scalar game, simultaneous GDA has update matrix $\mathbf{I}+\eta\mathbf{J}(\omega)$, whose eigenvalues are $1\pm i\eta\omega$ and have modulus
\[
\sqrt{1+\eta^2\omega^2}>1
\]
whenever $\eta\omega\neq0$. Therefore every nonzero scalar mode admits divergent trajectories. By Theorem~\ref{thm:transfer}, any network game containing a nonzero canonical frequency admits an initialization for which GDA diverges. Scalar games can therefore certify failure as well as success.

\paragraph{Time-average arguments and low-dimensional geometry.}
The scalar proof does not need to be spectral. In fact, low-dimensional geometry has already enabled time-average results that are difficult to obtain from generic online-learning bounds. \citet{bailey2019fast} prove $\Theta(\sqrt{T})$ regret for fixed-step gradient descent in two-agent, two-strategy zero-sum games by developing a tight geometric description of the non-equilibrating trajectories; time-average convergence follows as a corollary. That normal-form result is outside the exact unconstrained setting of Theorem~\ref{thm:transfer}, but it illustrates the methodological point: once the relevant trajectory geometry is genuinely low dimensional, optimization and averaging guarantees can become much easier to establish. In our setting, any scalar time-average rate or Nash-residual bound obtained in this way transfers through the same argument as the last-iterate result.

\subsection{Geometric interpretation and visualization}
\label{sec:visualization}

The same canonical coordinates also admit a useful geometric interpretation. In the disc-game literature, each $2\times2$ block is viewed as a planar competitive game in which the response field is a ninety-degree rotation and cyclic advantage is encoded geometrically. These coordinates have been used for visualization, strategic feature extraction, and low-rank summaries of game structure~\citep{balduzzi2018reevaluating,strang2024pta}. Recent work uses the same geometry to interpret learning and selection trajectories~\citep{lechon2025disc}. We do not claim these visualization ideas as a contribution; they matter here because the scalar systems that suffice analytically are also simple enough to inspect directly.

\subsection{Canonical modes as a route to acceleration}
\label{sec:acceleration}

The transfer principle suggests an analogy with classical acceleration for positive-definite quadratic optimization. Consider
\[
\min_{\mathbf{x}}\frac12\mathbf{x}^\top\mathbf{H}\mathbf{x},
\qquad
\mathbf{H}=\mathbf{Q}\mathbf{\Lambda}\mathbf{Q}^\top,
\qquad
0<\mu\mathbf{I}\preceq\mathbf{H}\preceq L\mathbf{I}.
\]
After the orthogonal change of coordinates $\mathbf{z}=\mathbf{Q}^\top\mathbf{x}$, gradient descent separates into scalar recurrences
\[
z_{i,t+1}=(1-\eta\lambda_i)z_{i,t}.
\]
This reduction turns high-dimensional algorithm design into a one-dimensional approximation problem over the spectral interval $[\mu,L]$. Chebyshev semi-iterative methods, Polyak momentum, and Nesterov acceleration all exploit this mode-wise viewpoint in different forms~\citep{golubVarga1961chebyshev,polyak1964speeding,nesterov2004introductory}. Related work has also used spectral-shape manipulation to accelerate smooth game dynamics~\citep{azizian2020accelerating}.

The analogous elementary object in a zero-sum game is the planar system
\[
\mathbf{J}(\omega)=
\begin{bmatrix}
0&\omega\\
-\omega&0
\end{bmatrix}.
\]
Thus a candidate accelerated learning rule can be designed and tuned by analyzing how it behaves as $\omega$ varies over the relevant frequency range. The geometry is particularly useful because each mode can be drawn directly: damping, rotation, phase error, and transient behavior are visible in two dimensions. The transfer theorem then converts a guarantee proved on the scalar games into the corresponding network guarantee. We view this as a promising route for accelerated game-learning methods, although developing such methods is outside the scope of this paper.

\subsection{Implications for numerical testing}
\label{sec:numerical-testing}

The exact worst-mode identity in Corollary~\ref{cor:worst-mode} also simplifies numerical evaluation. For a network game with canonical frequencies $\omega_1,\ldots,\omega_\ell$, the worst-case finite-time response is
\[
    R_T(\eta,\mathbf{G})
    =
    \max_{1\leq j\leq\ell} r_T(\eta,\omega_j).
\]
Thus, once the scalar response $r_T(\eta,\omega)$ has been evaluated, there is no need to simulate the full network to determine its worst-case Euclidean response. One need only evaluate the scalar response at the canonical frequencies of the game. More broadly, when comparing a learning rule over a spectral family $\Omega$, the corresponding worst-case numerical problem reduces to
\[
    \sup_{\omega\in\Omega} r_T(\eta,\omega).
\]
This reduction is exact for the performance criteria covered by Corollary~\ref{cor:worst-mode}, rather than an empirical approximation. It also turns learning-rate tuning over such a family into the low-dimensional minimax problem
\[
    \min_{\eta}\;\sup_{\omega\in\Omega} r_T(\eta,\omega),
\]
providing a direct numerical counterpart to the mode-wise design perspective in Section~\ref{sec:acceleration}.

\subsection{Relationship to spectral analysis: SVD versus Schur}
\label{sec:svd-schur}

The scalar-game viewpoint is closely related to conventional spectral analysis. In a two-player game $\mathbf{x}_1^\top\mathbf{A}\mathbf{x}_2$, the singular value decomposition $\mathbf{A}=\mathbf{U}\mathbf{\Sigma}\mathbf{V}^\top$ is often the most natural representation because the two players can be transformed separately. This is one reason singular values appear so naturally in last-iterate analyses.

In the stacked network notation, the real Schur form is especially convenient because a single orthogonal matrix acts on the full state:
\[
\mathbf{Q}^\top\mathbf{G}\mathbf{Q}=\mathbf{J}.
\]
If $\tilde{\mathbf{x}}=\mathbf{Q}^\top\mathbf{x}$ and $\tilde{\mathbf{y}}=\mathbf{Q}^\top\mathbf{y}$, then
\[
\mathbf{x}^\top\mathbf{G}\mathbf{y}
=
\tilde{\mathbf{x}}^\top\mathbf{J}\tilde{\mathbf{y}}.
\]
Thus the same coordinate transformation appears on both sides of the interaction, and the gradient update inherits the planar block structure immediately. In the standard two-player case, SVD and Schur expose the same frequencies; we use Schur because it aligns directly with the stacked skew-symmetric operator and the planar games, not because it contains fundamentally different spectral information. The connection between these views is also discussed explicitly in PTA~\citep{strang2024pta}.

\subsection{Relation to constrained normal-form games}
\label{sec:constrained}

The global transfer theorem does not directly apply to simplex-constrained normal-form games: projections, multiplicative updates, and changes in support are not preserved by a generic orthogonal transformation. Nevertheless, several analyses of constrained learning already separate \emph{support identification or boundary behavior} from the subsequent dynamics within a fixed face of the simplex. For MWU, \citet{Bailey18Divergence} show that when no interior equilibrium exists, the dynamics approach the minimal product of subsimplices containing all Nash equilibria and then analyze the behavior within that reduced support. For optimistic MWU, \citet{daskalakis2019last} similarly separate the global approach to a neighborhood of equilibrium from a local contraction argument. More recently, \citet{nan20251} prove an $O(1/T)$ rate for alternating GDA when an interior equilibrium exists and a corresponding local $O(1/T)$ guarantee in the non-interior case.

These results suggest a possible role for the present decomposition beyond the unconstrained model. Once a support is identified and the dynamics remain in the relative interior of a fixed face, the underlying payoff field is bilinear on an affine subspace; this makes low-dimensional modal analysis a natural candidate for clarifying the local within-support behavior of gradient-type dynamics. The update geometry of projected, multiplicative, or regularized methods can still differ substantially from the unconstrained recurrence studied here, so no constrained transfer theorem is claimed. The difficult transient question of how the support is found remains separate. We view the separation---support identification first, low-dimensional local analysis second---as a promising route toward constrained extensions.

\section{Conclusion}
\label{sec:conclusion}

Low-dimensional decompositions of skew-symmetric games are classical and have already been used for spectral analysis, visualization, and the study of particular learning rules. The contribution of this paper is to make their learning-theoretic implication explicit: for the gradient-based update family considered here, a convergence theorem on scalar zero-sum games can be transferred directly to arbitrary unconstrained network zero-sum games.

The resulting principle is simple. Scalar convergence conditions determine network learning-rate conditions; the exact worst-case network rate equals the largest scalar canonical rate; and scalar time-average arguments can be lifted even when their proofs are geometric rather than spectral. In this precise sense, scalar zero-sum games form a complete test family for universal convergence guarantees of the update class studied here. This suggests treating scalar games as standard test problems for analyzing and comparing gradient-based learning rules. It also points toward a design methodology analogous to classical acceleration in quadratic optimization: understand and optimize the rule on the canonical low-dimensional games, then transfer the resulting guarantee to the network game.

The scope of the result is also now clear. The theorem applies exactly to unconstrained bilinear network games and the symmetric gradient-based recurrences studied here; affine costs and fixed positive player-specific learning-rate multipliers admit closely related reductions. Constrained normal-form games require a separate support-identification phase before local bilinear structure can be exploited. More generally, extending the transfer principle to genuinely asymmetric update rules, multiplicative-weights or other FTRL dynamics, nonlinear games, and asynchronous learning remains open. These questions concern the boundary of the transfer principle itself and are natural directions for future work.

\section*{Generative AI Use Statement}

The initial paper outline and proofs were developed entirely by the authors. Generative AI tools (OpenAI ChatGPT 5.6) were subsequently used to assist with drafting and revising text, improving exposition and organization, checking notation and internal consistency, and providing reviewer-style feedback. The tool was also used as an interactive aid in discussing the presentation and formulation of mathematical arguments and in identifying potentially relevant literature. All mathematical results, proofs, interpretations, and citations were independently reviewed and verified by the authors. The authors made all substantive research decisions and take full responsibility for the content of the manuscript.


\bibliographystyle{plainnat}
\bibliography{references}

\end{document}